\documentclass[amssymb, floatfix, superscriptaddress, aps, prb, reprint, longbibliography]{revtex4-1}
\usepackage{graphicx}
\usepackage{dcolumn}
\usepackage{color}
\usepackage{physics}
\usepackage{amsmath}  
\usepackage{hyperref}
\usepackage{braket}
\hypersetup{colorlinks=true,citecolor=magenta,linkcolor=magenta, urlcolor=magenta}

\def\up{\uparrow}
\def\dn{\downarrow} 
\def\br{\mathbf{r}}
\def\bx{\mathbf{x}}
\def\exc{E_{\rm XC}}
\def\ex{E_{\rm X}}

\newcommand{\pr}[2]{ \ket{#1}\!\bra{#2} }
\newcommand{\hc}{\mathop{\text{h.c.}}}

\bibpunct{[}{]}{,}{n}{}{} 

\begin{document}
\setlength{\abovedisplayskip}{7pt}
\setlength{\belowdisplayskip}{7pt}

\title{Physical Origin of the One-Quarter Exact Exchange in Density Functional Theory}  

\author{Marco Bernardi}
\email{bmarco@caltech.edu}
\affiliation {Department of Applied Physics and Materials Science, California Institute of Technology, Pasadena, California 91125}

\begin{abstract} 
Exchange interactions are a manifestation of the quantum mechanical nature of the electrons and play a key role in predicting the properties of materials from first principles. 
In density functional theory (DFT), a widely used approximation to the exchange energy combines fractions of density-based and Hartree-Fock (exact) exchange. 
This so-called hybrid DFT scheme is accurate in many materials, for reasons that are not fully understood. 
Here we show that a 1/4 fraction of exact exchange plus a 3/4 fraction of density-based exchange is compatible with a correct quantum mechanical treatment of the exchange energy of an electron pair in the unpolarized electron gas.   
We also show that the 1/4 exact-exchange fraction mimics a correlation interaction between doubly-excited electronic configurations.  
The relation between our results and trends observed in hybrid DFT calculations is discussed, along with other implications. 
\end{abstract}

\maketitle
In the framework of density functional theory (DFT), 
the total energy can be expressed as a functional of the electron density~\cite{HK}, 
and minimized to obtain the ground state~\cite{kohn1965}. 
As the electron interactions cannot at present be expressed exactly in terms of the density, 
one needs to approximate the exchange and correlation energies, often using a local functional of the density as is done in the local density approximation (LDA) 
and generalized gradient approximations (GGAs) of DFT~\cite{LDA,GGA}. 
So-called hybrid exchange-correlation functionals~\cite{Becke-1993} are a popular choice in modern DFT calculations. They replace part of the density-based exchange energy with Hartree-Fock (HF) 
(so-called \lq\lq exact\rq\rq) exchange, improving the accuracy of several computed properties.\\  
\indent
In the simplest hybrid functional, called PBE0~\cite{Becke-PBE0, Adamo, Scuseria}, a single parameter $\alpha$ equal to the fraction of HF exchange 
is employed for the exchange mixing, according to $\exc = \exc^{\rm DFT} + \alpha\, (\ex^{\rm HF} - \ex^{\rm DFT})$, where $E_{\rm X}$ and $\exc$ are  
the exchange and exchange-correlation energies, respectively, and the DFT and HF superscripts denote density-based and HF-based quantities.  
It has been shown that an exact-exchange fraction of approximately $\alpha\!=\!1/4$ gives optimal ground state energies and properties~\cite{Perdew-rationale}, 
especially when used in conjunction with an accurate GGA functional.  
In the more advanced range-separated HSE hybrid functional~\cite{HSE}, which separates the short- and long-range parts of exchange, an optimal fraction of short-range exchange close to 1/4 has also been found.  
At present, the 1/4 fraction is an empirical parameter without a rigorous justification.  
Heuristic explanations exist for the physical origin of exact exchange mixing and why it improves the ground state energy~\cite{Becke-PBE0,Perdew-rationale}, 
but rigorous results on this topic are still scarce.
\\
\indent
%
%
Here we show that a hybrid functional with 1/4 exact exchange mixing mimics the correct quantum mechanical (QM) treatment of the exchange energy of an electron pair in an unpolarized electron gas. 
Our analysis compares a statistical ensemble with a pure-state superposition of spin states of an electron pair, showing that a 3/4 fraction of density-based exchange reflects the ratio of triplet to total spin states of the electron pair, 
while the 1/4 exact exchange is associated with a spin-flip exchange interaction between doubly-excited configurations, which contributes to the ground state energy. 
Taken together, these results link exact-exchange mixing with the treatment of electron pair spin states in DFT.  
\vspace{-10pt}
\section*{Results}
\begin{figure}[t!]
\includegraphics[width=0.85\columnwidth]{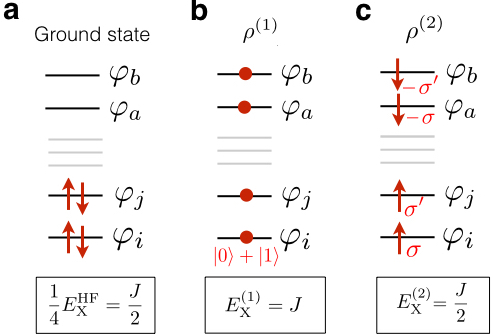} 
\caption{Electronic configurations considered in this work.  
(a) Ground state with doubly-occupied orbitals $\varphi_i$ and $\varphi_j$.  
(b) Configuration used to derive $\rho^{(1)}$, in which the orbitals $\varphi_i$ and $\varphi_j$ are both occupied by a single electron with spin state $(\, \ket{0} + \ket{1} \,)/\sqrt{2}$. 
(c) Configurations used to derive $\rho^{(2)}$, with orbitals $\varphi_i$ and $\varphi_j$ each occupied by one electron, with spin $\sigma$ and $\sigma'$, respectively.  
In (b) and (c), an electron has been promoted from each of the orbitals $\varphi_i$ and $\varphi_j$ to  $\varphi_a$ and $\varphi_b$, in a way that conserves the spin unpolarized character.  
}\label{fig1}
\end{figure}

\vspace{-10pt}
\noindent
\textbf{Exchange interactions.} 
The exchange interaction lowers the electron repulsion energy by keeping pairs of electrons apart. 
Though exact in principle, DFT is a QM theory based on a real-valued \textit{probability} $-$ the electron density $-$ rather than a complex-valued probability amplitude, as is the wave function. 
Treating exchange is challenging for a local QM theory based on probability such as DFT, 
given that exchange is an inherently nonlocal interaction written in terms of electron orbitals. 
In HF, a pairwise treatment of exchange is adopted, in which only particles occupying spin-orbitals with parallel spin contribute to the exchange energy. 
The density-based picture of DFT is rather different $-$ exchange is not treated as a sum of pairwise interactions, but rather, it depends only on the electron density at a point. 
This means that all particles in the density contribute equally to exchange in semilocal DFT, regardless of their spin state. 
The main challenge to understanding exact-exchange mixing is marrying these two widely different approaches to treat exchange.
\\
\indent
We focus on the $N$-particle interacting electron gas in a material, and assume it is unpolarized with a ground state [see Fig.~\ref{fig1}(a)] consisting of $N$/2 doubly-occupied orthonormal orbitals, $\varphi_i$ ($i=1, ...\,, N/2$). 
Its HF exchange energy is:
\begin{equation}
E^{\rm HF}_{\rm X} = \frac{1}{2} \sum_{i,j=1}^{N/2} \,( 2 J )
\label{HF}
\end{equation}
where $J$ is the exchange interaction between orbitals $\varphi_{i}$ and $\varphi_{j}$~\cite{Grosso},
\begin{equation}
J = - \,e^2\! \int d\br_1 d\br_2  \frac{\varphi_i^*(\br_1) \varphi_j^*(\br_2) \varphi_j(\br_1) \varphi_i(\br_2)}{|\br_1 - \br_2 |}, 
\label{exchange}
\end{equation}
and the factor of 2 multiplying $J$ in Eq.~(\ref{HF}) accounts for the sum over spin. 
According to Eq.~(\ref{HF}), the total HF exchange energy is the sum of the exchange interaction between all orbital pairs, with each orbital pair contributing $2 J$ to the total exchange energy.\\
\indent 
When the same system is treated in DFT, the exchange energy is a functional of the electron density, $n(\br) = 2\sum_i |\varphi_i (\br)|^2$, 
but its expression is unknown and needs to be approximated. 
In the LDA~\cite{LDA}, for example, $E_{\rm X}^{\rm DFT}$ is approximated as a sum of local contributions from a homogeneous electron gas (HEG), 
while in the PBE0 hybrid functional  
with $1/4$ mixing, the exchange energy is
\begin{equation}
 E_{\rm X}^{\rm hybrid} = \frac{3}{4} E^{\rm DFT}_{\rm X} + \frac{1}{4} E^{\rm HF}_{\rm X}.
\end{equation}
\indent  
We focus on two doubly-occupied orbitals $\varphi_i$ and $\varphi_j$ composing the ground state [see Fig.~\ref{fig1}(a)],  
whose contribution to the HF exchange energy is $E^{\rm HF}_{\rm X} = 2 J$ 
(one quarter of their contribution is $J/2$). 
When $\varphi_i$ and $\varphi_j$ are the only two orbitals contributing to the charge density, the hybrid DFT exchange energy becomes 
\begin{equation}
E_{\rm X}^{\rm hybrid} = \frac{3}{4} E^{\rm DFT}_{\rm X}[n] + \frac{J}{2}, 
\label{pair}
\end{equation} 
where $n(\br) = 2\, (\, |\varphi_i(\br)|^2 + |\varphi_j(\br)|^2\,)$ is the density from the four electrons occupying the two orbitals $\varphi_i$ and $\varphi_j$.\\

\noindent %
\textbf{Models of an electron pair.} We model an electron pair in an unpolarized electron gas, 
writing its Coulomb repulsion energy $E_{\rm p}$ as a functional of its two-particle density matrix, 
$\rho(\bx_1,\bx_2)$, using~\cite{Lowdin,Coleman} 
\begin{equation}
E_{\rm p} = \int d\bx_1 d\bx_2\, V_{12}(\br_1,\br_2)\, \rho(\bx_1,\bx_2),
\label{erho} 
\end{equation}
where $V_{12} \!=\! \frac{e^2}{|\br_1 - \br_2|}$ is the two-body Coulomb interaction.\\
\indent
We model the pair as made up by an electron in orbital $\varphi_i$ with spin $\sigma$ and an electron in orbital $\varphi_j$ with spin $\sigma'$, 
and investigate its two-particle density matrix and exchange energy. If we measure the spin of the two electrons composing the pair, 
as the electron gas is unpolarized we expect to obtain one of these \textit{four} outcomes with equal probability: spin up for both electrons, spin down for both electrons, 
spin up for the electron in orbital $\varphi_i$ and spin down for the electron in orbital $\varphi_j$, or viceversa. 
%
%
We express each of these four scenarios, respectively, with a two-particle Slater determinant formed with spin-orbitals $\varphi_{i\sigma}$ and $\varphi_{j\sigma'}$ and properly normalized: 
\begin{align}
\begin{split}
	&\ket{\Psi_1} \equiv \ket{\up\up} =  \det[\varphi_{i\up},\varphi_{j\up}] \\ 
	&\ket{\Psi_2} \equiv \ket{\dn\dn} =  \det[\varphi_{i\dn},\varphi_{j\dn}] \\ 
	&\ket{\Psi_3} \equiv \ket{\up\dn} =  \det[\varphi_{i\up},\varphi_{j\dn}] \\ 
	&\ket{\Psi_4} \equiv \ket{\dn\up} =  \det[\varphi_{i\dn},\varphi_{j\up}].
\label{basis}
\end{split}
\end{align} 
Explicit expressions for these states are, for example, 
\begin{align}
\begin{split}
\Psi_1(\bx_1,\bx_2) &=\frac{ \varphi_{i} (\br_1) \varphi_{j}(\br_2) - \varphi_{j} (\br_1) \varphi_{i}(\br_2) }{\sqrt{2}} \times \ket{00} \\
\Psi_2(\bx_1,\bx_2) &=\frac{ \varphi_{i} (\br_1) \varphi_{j}(\br_2) - \varphi_{j} (\br_1) \varphi_{i}(\br_2) }{\sqrt{2}}  \times \ket{11},
\label{anti}
\end{split}
\end{align}
and similar ones for $\ket{\Psi_3}$ and $\ket{\Psi_4}$, 
%
%
%
%
\begin{figure*}[t!]
\includegraphics[width=1.9\columnwidth]{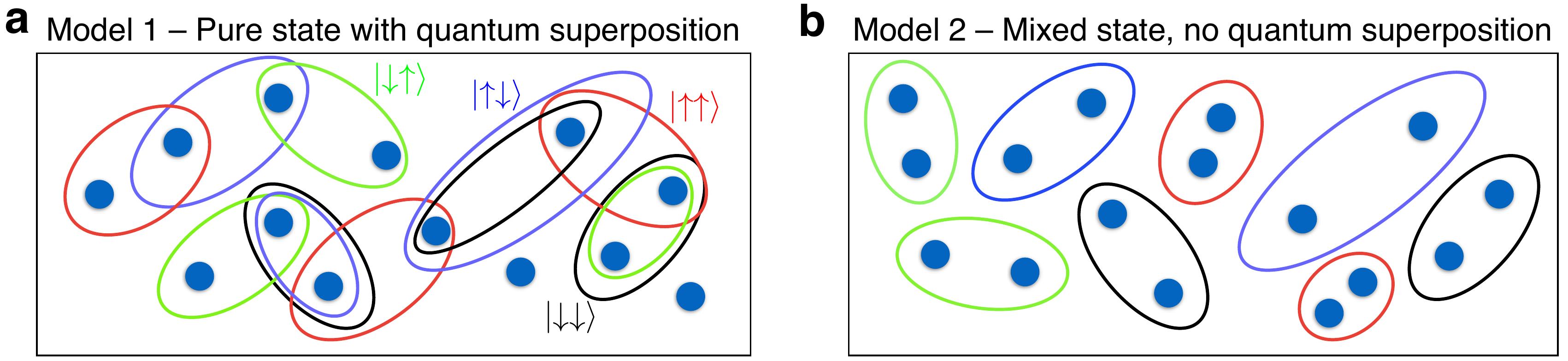}
\caption{Pictorial view of the two models employed to describe an electron pair in this work. (a) Model 1, a pure state, is the simplest \textit{ansatz} that includes quantum superposition and two-body correlations. 
(b) Model 2, a mixed state consisting of a statistical ensemble composed in equal parts by the four spin states of the pair, is a model based on probability, and thus similar in spirit to DFT. 
The colors label the pair basis states, $\ket{\Psi_\mu}$ in Eq.~(\ref{basis}).
}\label{fig2}
\end{figure*}
where $\ket{0}$ and $\ket{1}$ represent spin up and spin down states, respectively, 
and the tensor product $\ket{\sigma\sigma'} = \ket{\sigma} \otimes \ket{\sigma'}$ is a spin state of the two electrons~\cite{Nielsen}. 
\\
\indent  
Using the states in Eq.~(\ref{basis}) as a basis set, we formulate two models of the electron pair, both consistent with the spin measurements discussed above. 
The first approach, called here \mbox{\textit{model 1}}, describes the pair with a wave function $\ket{\Psi_{\rm p}}$ consisting of an equal quantum superposition of the four states:
\begin{equation}
\ket{\Psi_{\rm p}} = \frac{1}{\sqrt{4}}(\ket{\up\up} + \ket{\dn\dn} + \ket{\up\dn} + \ket{\dn\up}).
\end{equation}
The corresponding two-particle density matrix, $\rho^{(1)}$, is a so-called \textit{pure state}~\cite{Nielsen}, 
\begin{equation}
\rho^{(1)} = \ket{\Psi_{\rm p}}\!\bra{\Psi_{\rm p}} = \frac{1}{4}\,\sum_{\mu,\nu=1}^4 \pr{\Psi_\mu}{\Psi_{\nu}}.
\label{def:rho1}
\end{equation}
\noindent
This description of the electron pair is the simplest \textit{ansatz} that includes quantum superposition and two-body correlation effects.
\\
\indent
We contrast this description of the pair with one based on probability alone, called here \textit{model 2}, which is similar in spirit to DFT. 
In model 2, the pair is described as a statistical ensemble composed in equal parts by the four pair basis states in Eq.~(\ref{basis}).  
This model corresponds to a \textit{mixed state}~\cite{Nielsen}, generally described by the density matrix $\rho = \sum_{\mu} P_{\mu} \pr{\Psi_\mu}{\Psi_\mu}$. 
For our \mbox{model 2}, the two-particle density matrix is 
\begin{equation}
    	\rho^{(2)} = \frac{1}{4} \sum_{\mu=1}^4 \ket{\Psi_\mu}\!\bra{\Psi_\mu},
\end{equation}
so that in our case $P_\mu = 1/4 $ is the probability of finding the electron pair in state $\ket{\Psi_{\mu}}$. 
The two models of the pair, and their associated density matrices $\rho^{(1)}$ and $\rho^{(2)}$, are represented schematically in Fig.~\ref{fig2}. 
Spin measurements on the electron pair will give, both in model 1 and 2 (but for different reasons), one of the four states in Eq.~(\ref{basis}) with equal probability of 1/4, 
consistent with our assumption of an unpolarized electron gas.\\
\indent
%
%
An important property of $\rho^{(1)}$ and $\rho^{(2)}$ is that they are $N$-representable~\cite{Coleman}, 
in the sense that they can both be derived from two-body reduced density matrices, each obtained from an $N$-body density matrix by tracing out $N-2$ electrons~\cite{Coleman}. 
The $N$-representability guarantees that the energy derived using Eq.~(\ref{erho}) is physically meaningful~\cite{Coleman}; 
it further reveals the origin of $\rho^{(1)}$ and $\rho^{(2)}$ in terms of many-electron configurations. 
As we show in the Appendix, one can derive $\rho^{(1)}$ and $\rho^{(2)}$ from doubly-excited configurations described by Slater determinants 
in which an electron has been removed from each of the orbitals $\varphi_i$ and $\varphi_j$ and placed into unoccupied states [see Fig.~\ref{fig1}(b)$-$(c)]. 
Due to Brillouin's theorem, and as is known in configuration interaction theories,  
such doubly-excited configurations contribute to the ground state energy~\cite{Szabo,CC}.
\\ 

\begin{figure}[t!]
\includegraphics[width=0.95\columnwidth]{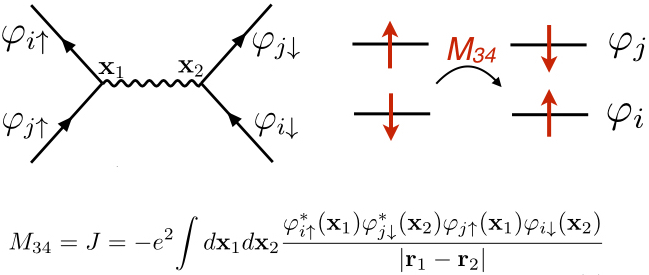}
\caption{Feynman diagram (left) and schematic visualization (right) of the spin-flip exchange interaction $M_{34}$ coupling the states $\ket{\Psi_3} = \ket{\up\dn}$ and $\ket{\Psi_4}=\ket{\dn\up}$. 
This interaction can only occur between doubly-excited configurations in which the orbitals $\varphi_i$ and $\varphi_j$ are each occupied by only one electron.
}\label{fig3}
\end{figure}

\noindent
\textbf{One-quarter exact exchange.}  
To obtain the exchange energy of the electron pair in the two models, $E^{(i)}_{\rm X}$ ($i=1,2$), 
we evaluate Eq.~(\ref{erho}) by computing the matrix elements of the Coulomb interaction $V_{12}$ between the two-particle determinant basis states in Eq.~(\ref{basis}), and keep only the exchange part of the result:
\begin{align}
\begin{split}
	&E^{(1)}_{\rm X} = \Tr[\rho^{(1)}V_{12}]_{\rm X} = \frac{1}{4}\sum_{\mu \nu} \mel{\Psi_{\mu}}{V_{12}}{\Psi_{\nu}}_{\rm X} = \frac{1}{4}\sum_{\mu \nu} M_{\mu \nu} \\
	&E^{(2)}_{\rm X} = \Tr[\rho^{(2)}V_{12}]_{\rm X} = \frac{1}{4}\sum_{\mu} \mel{\Psi_{\mu}}{V_{12}}{\Psi_{\mu}}_{\rm X} = \frac{1}{4}\sum_{\mu} M_{\mu \mu},
\label{energies}
\end{split}
\end{align}
where the sums run over the four basis states and the exchange part (subscript X) of the matrix elements is defined as
\begin{equation}
M_{\mu\nu} \equiv \braket{\Psi_{\mu}| V_{12} |\Psi_{\nu}}_{\rm X}.
\end{equation} 
With these definitions, the exchange matrix $M$ becomes (see Appendix):
\begin{align}
\label{mmatrix}
M = 
         \begin{pmatrix}
    	J &  0 &  0 &  0 \\
    	0 & J  & 0  &  0 \\
    	 0 & 0   & 0 &  J \\
    	 0 &  0  &  J  & 0 
    	\end{pmatrix} \,.
\end{align}
%
The only nonzero off-diagonal matrix element is \mbox{$M_{34} = \braket{\up\dn | V_{12} | \dn\up}_{\rm X} = J$} (and $M_{43} \!=\! M_{34}$), 
which corresponds to a spin-flip exchange interaction (see Fig.~\ref{fig3}) between pair states with antiparallel spins in the orbitals $\varphi_i$ and $\varphi_j$. 
This interaction contributes to the exchange energy only in model 1 [see Eq.~(\ref{energies})]  
and is associated with an exchange process between two doubly-excited configurations  
in which an electron in spin-orbital $\varphi_{j\up}$ scatters into $\varphi_{i\up}$, while an electron in spin-orbital $\varphi_{i\dn}$ scatters into $\varphi_{j\dn}$, due to the Coulomb interaction (see Fig.~\ref{fig3}). 
As this interaction is not included in HF, it should be regarded as a correlation interaction $-$ in the guise of an exchange process $-$ contributing to the ground state energy. 
\\
\indent 
%
%
The exchange energy of the pair in the two models is obtained from Eq.~(\ref{energies}) with the matrix $M$ given above:
\begin{align}   
\begin{split}
	E^{(1)}_{\rm X} &= \frac{1}{4}\sum_{\mu \nu} M_{\mu \nu} = J \\
	E^{(2)}_{\rm X} &= \frac{1}{4}\sum_{\mu} M_{\mu \mu} = \frac{J}{2}.
\end{split}
\end{align}
It is noteworthy that the QM treatment of model 1 accounts for all four processes contributing to exchange, including the parallel-spin exchange processes, $M_{11}$ and $M_{22}$, 
and the two spin-flip exchange processes $M_{34}$ and $M_{43}$.  
By contrast, the probability-based ensemble description of the pair in model 2 misses the two spin-flip exchange process $M_{34}$ and $M_{43}$. 
\indent
As model 1 is a proper QM treatment of the pair, we rename $E^{(1)}_{\rm X}$ to just $E_{\rm X}$, the exchange energy of the electron pair, and write:
\begin{equation}
E_{\rm X} = E^{(2)}_{\rm X} + \frac{J}{2}.
\label{result-part}
\end{equation}
%
The last term in Eq.~(\ref{result-part}) equals one quarter of the HF exchange energy contribution from the interaction between the two orbitals $\varphi_i$ and $\varphi_j$, so we can write:
\begin{equation}
E_{\rm X} = E^{(2)}_{\rm X} + \frac{1}{4} E^{\rm HF}_{\rm X}.
\label{result1}
\end{equation}
%
%

\noindent
\textbf{Combining density-based exchange.} 
We wish to express the exchange energy in the statistical ensemble of model 2, $E^{(2)}_{\rm X}$ in Eq.~(\ref{result1}), in the density-based picture. 
We show below that $E^{(2)}_{\rm X}$ can be approximated as 3/4 of the density-based exchange energy  
and thus the exchange energy for a pair of occupied orbitals is
\begin{equation}
E_{\rm X} = \frac{3}{4} E^{\rm DFT}_{\rm X}[n] + \frac{1}{4} E^{\rm HF}_{\rm X}.
\label{final}
\end{equation}
Comparing this result with Eq.~(\ref{pair}) tells us that 1/4 exact-exchange mixing provides an accurate description of the exchange interaction of an electron pair, 
accounting for both quantum superposition and two-body correlation effects.  
\\
\indent
Writing $E^{(2)}_{\rm X}$ in terms of the density requires shifting from a pairwise to a density-based view of exchange interactions. 
To express exchange as a local interaction depending explicitly on the electron density (and thus, on the electron pair density), 
we assume that only electron pairs in a \textit{triplet} state contribute to the exchange energy at a point due to their spatially antisymmetric wave function. 
%
%
\begin{figure*}[t!]
\includegraphics[width=1.95\columnwidth]{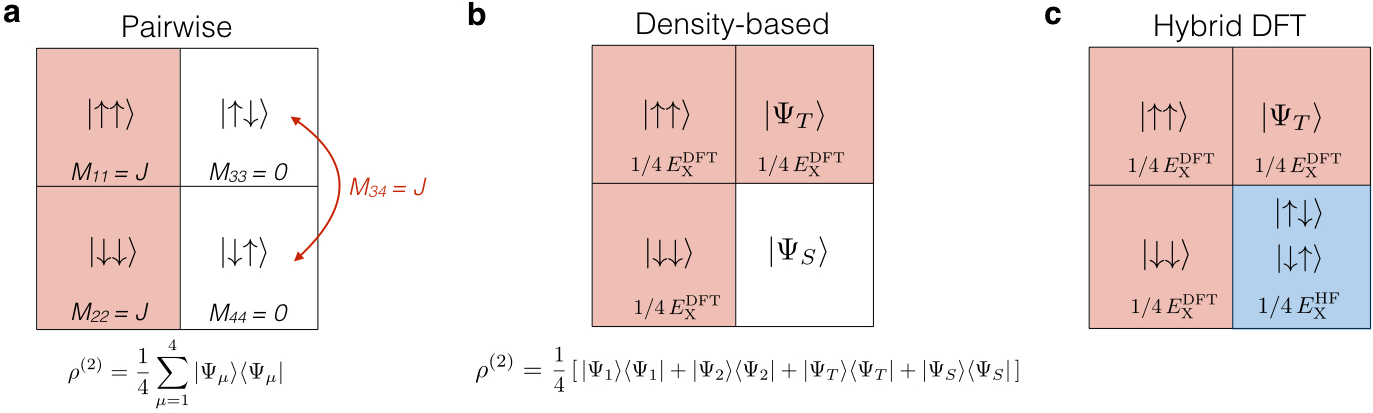}
\caption{Two schematic views of the electron gas in model 2 are given in (a) and (b). In both panels, only the portions in red contribute to exchange. 
(a) The pairwise view, in which $\rho^{(2)}$ is an ensemble of states with well-defined single particle spin, but one is unsure of how to account for the spin-flip exchange interaction $M_{34}$. 
(b) Our view of density-based exchange, in which $\rho^{(2)}$ is a mixture of 3 triplet and 1 singlet states. Each triplet spin state contributes $1/4\, E_{\rm X}^{\rm DFT}$ to exchange, 
so that a total of 3/4 of the density-based exchange interaction needs to be included. (c) Schematic representation of exchange in hybrid DFT with 1/4 mixing. 
The portions of the system in red contribute to density-based exchange; the quarter in blue is the exact exchange contribution, which accounts, as we argue, for spin-flip exchange processes, thus completing the exchange hole.
}\label{fig4}
\end{figure*}
In our model 2 of the electron pair, both $\ket{\Psi_1} = A_{ij}\ket{00}$ and $\ket{\Psi_2} = A_{ij}\ket{11}$ are spatially antisymmetric [see Eq.~(\ref{anti})] triplet states. 
However, the spatial parts of $\ket{\Psi_3} = \ket{\up\dn}$ and $\ket{\Psi_4} = \ket{\dn\up}$ lack a well-defined symmetry. 
The problem is clearer when shown in a pictorial view of the statistical ensemble $\rho^{(2)} = \frac{1}{4} \sum_\mu \pr{\Psi_\mu}{\Psi_\mu}$ [see Fig.~\ref{fig4}$\,$(a)], 
where the portions of the system in states $\ket{\Psi_1}$ and $\ket{\Psi_2}$ clearly contribute to exchange with their corresponding matrix elements $M_{11} = M_{22} = J$, 
but it is unclear how the other half of the system, in states $\ket{\Psi_3}$ and $\ket{\Psi_4}$, contributes, since the spin-flip exchange process hides in the interaction $M_{34}$ between these two states.\\
\indent
This subtle issue can be resolved by diagonalizing $M$ in the subspace of states $\ket{\Psi_3}$ and $\ket{\Psi_4}$, which provides a singlet and a triplet state with well-defined spatial symmetry. 
Such triplet state $\ket{\Psi_{\rm T}}$ and singlet state $\ket{\Psi_{\rm S}}$ can be written in terms of $\ket{\Psi_3}$ and $\ket{\Psi_4}$ as
\begin{align}
\begin{split}
\ket{\Psi_{\rm T}} &= \frac{1}{\sqrt{2}} \left(\, \ket{\Psi_3} + \ket{\Psi_4} \right)= A_{ij}(\br_1,\br_2) \times \left( \frac{ \ket{01} + \ket{10} }{\sqrt{2}} \right) \\
\ket{\Psi_{\rm S}} &= \frac{1}{\sqrt{2}} \left(\, \ket{\Psi_3} - \ket{\Psi_4} \right) = S_{ij}(\br_1,\br_2) \times \left( \frac{ \ket{01} - \ket{10} }{\sqrt{2}} \right)
\end{split}
\end{align}
where $A_{ij}(\br_1,\br_2) \!=\! [ \varphi_i(\br_1) \varphi_j(\br_2) - \varphi_j(\br_1) \varphi_i(\br_2) ]/\sqrt{2}$ is the antisymmetric spatial wave function used above, 
and $S_{ij}(\br_1,\br_2) \!=\! [ \varphi_i(\br_1) \varphi_j(\br_2) + \varphi_j(\br_1) \varphi_i(\br_2) ]/\sqrt{2}$ is a symmetric spatial wave function. 
It is clear that only $\ket{\Psi_{\rm T}}$ contributes to exchange at a given point due to its antisymmetric spatial wave function, but $\ket{\Psi_{\rm S}}$ does not.\\
\indent
The ensemble pair state $\rho^{(2)}$ can be rewritten as formed for 3/4 by triplet ($\ket{\Psi_1}$, $\ket{\Psi_2}$ and $\ket{\Psi_{\rm T}}$) and 1/4 by singlet ($\ket{\Psi_{\rm S}}$) spin states: 
\begin{align}
\begin{split}
\rho^{(2)} &= \frac{1}{4} \sum_{\mu=1}^4 \pr{\Psi_\mu}{\Psi_\mu} \\
               &= \frac{1}{4} \left(\, \pr{\Psi_1}{\Psi_1}  + \pr{\Psi_2}{\Psi_2} + \pr{\Psi_{\rm T}}{\Psi_{\rm T}} + \pr{\Psi_{\rm S}}{\Psi_{\rm S}} \,\right).
\end{split}              
\end{align}
This different way of expressing $\rho^{(2)}$, in terms of pair states with a well-defined total spin, is shown schematically in Fig.~\ref{fig4}$\,$(b). 
The part of the ensemble contributing to exchange, called below $\rho^{(2)}_{\rm E}$, contains the terms in $\rho^{(2)}$ proportional to $|A_{ij}(\br_1,\br_2)|^2$:
\begin{equation} 
\rho^{(2)}_{\rm E} = \frac{1}{4} \left(\, \pr{\Psi_1}{\Psi_1}  + \pr{\Psi_2}{\Psi_2} + \pr{\Psi_{\rm T}}{\Psi_{\rm T}} \,\right).
\end{equation}
We can thus show that the fraction of electrons $\alpha_{\rm E}$ contributing to the density-based exchange equals 3/4:  
\begin{equation} 
\alpha_{\rm E} = \frac{\int \!d\bx_1 n_{\rm E}(\bx_1)} {\int \!d\bx_1\, n(\bx_1)} = \frac{\int \!d\bx_1 d\bx_2\, \rho^{(2)}_{\rm E}(\bx_1,\bx_2)} {\int \!d\bx_1 d\bx_2\, \rho^{(2)}(\bx_1,\bx_2)} = \frac{3}{4}.
\end{equation}
Above, the electron density of the pair, $n(\bx)$, is obtained from the two-body density matrix as~\cite{Lowdin, Coleman} 
\begin{align}
\begin{split}
n(\bx) & = 2 \int\!\! d\bx_2\, \rho^{(2)}(\bx,\bx_2) \\
          & = \frac{1}{2} \left[\, |\varphi_i|^2(\br) + |\varphi_j|^2 (\br) \,\right] (\, \pr{0}{0} + \pr{1}{1} \,),
\end{split}
\end{align}
which correctly integrates to 2 electrons, 
while the electron density contributing to exchange is 
\begin{align}
\begin{split}
 n_{\rm E} (\bx) & = 2 \int\!\! d\bx_2\, \rho^{(2)}_{\rm E}(\bx,\bx_2) \\
          & \!\!= \frac{1}{2} \left[\, |\varphi_i|^2(\br) + |\varphi_j|^2 (\br) \,\right] \times \frac{3}{4}(\, \pr{0}{0} + \pr{1}{1} \,),
\end{split}
\end{align}
which integrates to $2 \times \frac{3}{4}$ electrons. 
As only the triplet states contribute to exchange,  
a fraction of $\alpha_{\rm E} = 3/4$ of each electron pair should be included in the density-based exchange energy, and thus we obtain $E^{(2)}_{\rm X} = 3/4\,\, E^{\rm DFT}_{\rm X}[n]$. 
This step concludes our proof of Eq.~(\ref{final}) and justifies Eq.~(\ref{pair}) for the hybrid DFT exchange energy contributed by a pair of electron orbitals. 
This result can be extended to the entire electron gas (see Appendix), justifying why hybrid DFT with 1/4 exchange mixing shows accurate results for spin-unpolarized systems in benchmark calculations.\\ 

\noindent
\textbf{Systems with less than four electrons.} 
We discuss numerical results that corroborate our interpretation of the origin of the 1/4 exact exchange, focusing on trends obtained with hybrid functionals in systems with less than four electrons. 
Our treatment of the physical origin of the 1/4 exact exchange is valid only when at least four electrons (in two doubly-occupied orbitals) are present, 
since otherwise $1/4\, E_{\rm X}^{\rm HF}$ does not equal the correlation processes due to doubly-excited configurations. 
A table of atomization energies of several molecules, taken from Ref.~\cite{Perdew-rationale}, shows that a simple PBE0 hybrid with 1/4 exact exchange dramatically improves 
the accuracy of the computed atomization energy compared to LDA or GGA results.  
The only three molecules in the data set with less than four electrons $-$ H$_2$, Li$_2$, and LiH $-$ are an exception to this trend~\cite{Perdew-rationale};    
their atomization energies computed with the hybrid functional exhibit nearly the same or a larger discrepancy with experiment than LDA or GGA. 
Table~\ref{table1} gives the atomization energies of H$_2$, Li$_2$ and LiH from Ref.~$\,$\onlinecite{Perdew-rationale}; 
they can be reproduced with any DFT package, as we have verified. 
It is noteworthy that these are the only three molecules in the data set with less than four valence electrons and also the only molecules for which the hybrid result does not improve over LDA or GGA. 
These trends support the interpretation put forward in this work for the physical origin of the 1/4 exact exchange.\\
\vspace{20pt}

\begin{table}[!h]
\vspace{-10pt}
\centering
		%
	\caption{Atomization energies (in eV units) taken from Ref.~\cite{Perdew-rationale} for molecules in which a hybrid with 1/4 exact exchange does not improve the accuracy of LDA and PW91 GGA result.\\}
	\label{table1}
	\begin{tabular}{ c l c c c } 
		\hline
		\hline
			\vspace{3pt}

	        Material & \,\,\, LDA & \,\,\,\,  GGA (PW91) & \,\,\,\,  1/4 Hybrid & \,\,\,\,  Experiment  \\
			 
		\hline
		\hline
		H$_2$  & \,\,\,\,\,113   &\,\,\,\,  105 &\,\,\,\, 105  & \,\,\,\, 109   \\ 
		LiH       &\,\,\,\,\,  60     &\,\,\,\,  53 &\,\,\,\, 52  &\,\,\,\,  58    \\ 
		Li$_2$ &\,\,\,\,\,  23     & \,\,\,\, 20 &\,\,\,\, 19 & \,\,\,\, 24  \\
		\hline\hline
	\end{tabular}
\end{table}

\section*{Discussion}
\vspace{-10pt}
Though this work focuses on the unpolarized electron gas, spin-polarized systems also provide valuable insight into exact-exchange mixing. 
In the fully spin-polarized limit, spin measurements on any electron pair would return a triplet state with certainty, so our models 1 and 2 become equivalent and the optimal amount of exact-exchange is zero. 
We thus expect semilocal DFT to describe a simple ferromagnetic metal better than hybrid DFT with 1/4 mixing.  
This observation is consistent with hybrid DFT results by Paier et al.~\cite{Paier}, who compared the performance of PBE and HSE for Fe and other itinerant magnetic systems. 
In Fe, the magnetic moment using the PBE functional is in excellent agreement with experiment, while HSE makes large errors on the magnetic moment and exchange splitting~\cite{Paier}. 
Another example of systems where 1/4 exact-exchange mixing fails are magnetic or correlated transition metal compounds, for which our results for an unpolarized electron gas do not apply. 
\\
\indent
Though our derivations focused on the PBE0 hybrid functional, widely used range-separated hybrid functionals, such as the HSE~\cite{HSE}, also employ a fraction of exact exchange close to 1/4 in their short-range exchange interactions. 
As correlation interactions are typically short-ranged, the success of the HSE functional in predicting the ground state supports the result shown here that the 1/4 exact exchange mimics correlations. 
In addition, it is known that the optimal $\sim$1/4 fraction of exact exchange examined here is appropriate when used in conjunction with GGA functionals (e.g., PBE or PW91), 
but not in general for mixing with the LDA exchange (although extensive tests for hybrid mixing with LDA are scarce). These trends in hybrid DFT calculations can be seen as a success of the GGAs  
as their optimal exact exchange mixing is close to 1/4, a value we predict to be physically justified. 
\\
\indent
Finally, this work investigates the ground state energy and does not examine the problem of band gap calculations using hybrid DFT, on which there has been recent progress~\cite{Marques,Skone}.
It has been shown that the optimal mixing parameter $\alpha$ for the band gap is roughly equal to the inverse static dielectric constant, $\alpha \approx 1/\epsilon_\infty$, and is thus system dependent~\cite{Marques,Skone}. 
Although hybrid functionals with $\alpha \approx 1/\epsilon_\infty$ mixing have also been used for structural properties~\cite{Galli-water}, 
their use for ground state calculations, to the author's knowledge, has not been rigorously justified. 
One could argue that, by admixing nonlocal exchange, hybrid functionals are flexible enough to mimic static screening for band gap  
and correlation for ground state calculations, for which a value of $\alpha \approx 1/4$ is appropriate in spin unpolarized systems as shown here.
\\
\indent
%
%
In summary, we presented a model of an electron pair that sheds light on the connection between exact-exchange mixing and the treatment of electron pair spin states in DFT. 
We argued that density-based exchange approximations regard the electron pair spin states as a statistical ensemble and can only capture triplet states. 
Our interpretation that the 1/4 exact exchange mimics correlations is consistent with observations from hybrid DFT calculations. 
Though this work does not propose an improved exchange-correlation functional, the hope is that the view of exchange interactions in DFT we put forward will stimulate new conceptual advances.
\section*{Acknowledgements}
\vspace{-10pt}
\noindent 
This work was supported by the National Science Foundation under Grant No. DMR-1750613.
\section*{Appendix}
\vspace{-10pt}
\noindent
\textbf{$N$-Representability of the Density Matrices $\rho^{(1,2)}$.} 
Our definitions given above for $\rho^{(1)}$ and $\rho^{(2)}$ include only the terms in the reduced density matrices contributing to the exchange interaction between orbitals $\varphi_i$ and $\varphi_j$. 
In this section, we show that one can obtain $\rho^{(1)}$ from the 2-body reduced density matrix of a pure state $\pr{ \Psi_N^{(1)} }{ \Psi_N^{(1)} }$, 
with $\ket{ \Psi_N^{(1)} }\!=\! \det[ \varphi_1,...\,, \varphi_{i}\!\ket{s}, ...\,, \varphi_{j}\!\ket{s}, ...\,] $ 
a Slater determinant in which each electron occupying $\varphi_i$ and $\varphi_j$ is in the spin superposition state \mbox{$\ket{s}\!=\!(\ket{0} \!+\! \ket{1})/\sqrt{2}$} 
[see Fig.~\ref{fig1}(b)].   
Similarly, $\rho^{(2)}$ can be obtained as the equal-probability ensemble of four density matrices, $\Gamma_{\sigma\sigma'}$, 
each obtained from the 2-body reduced density matrix of a pure state $\pr{ \Psi_{N,\sigma\sigma'}^{(2)} }{ \Psi_{N,\sigma\sigma'}^{(2)} }$ with Slater determinant 
$\ket{\Psi_{N,\sigma\sigma'}^{(2)} } = \det[ \varphi_1,...\,, \varphi_{i\sigma}, ...\,, \varphi_{j\sigma}, ...\,] $ [see Fig.~\ref{fig1}(c)].\\
\indent
Following Lowdin~\cite{Lowdin}, we define the 2-body reduced density matrix (2-RMD) $\widetilde{\Gamma}(\bx_1,\bx_2)$ for $N$ electrons as
\begin{equation}
\widetilde{\Gamma}(\bx_1,\bx_2) = {N \choose 2} \int\! d\bx_3 ...\, d\bx_N |\Psi_N (\bx_1,\bx_2,\bx_3,...\,,\bx_N) |^2 
\label{app:gamma}
\end{equation}
where $\Psi_N (\bx_1,\bx_2,\bx_3,...\,,\bx_N)$ is an antisymmetric wave function for $N$ electrons, and $\bx=\{ \br,\sigma \}$ denotes both the spatial coordinate $\br$ and the spin $\sigma$.\\
\indent
We specialize to the case in which the wave function $\Psi_N$ is a single Slater determinant, $\Psi_N = (1/\sqrt{N!}) \det[ \{ \varphi_k \}]$,  
with $\{ \varphi_k \}$ a set of orthonormal spin-orbitals labeled by the index $k=1,...\,,N$. In this case, the 2-RMD takes the simple form~\cite{Martin}
\begin{align}
\widetilde{\Gamma}(\bx_1,\bx_2) &= \frac{1}{2}
\begin{vmatrix}
\gamma(\bx_1,\bx_1) & \gamma(\bx_1,\bx_2)  \\ 
\gamma(\bx_2,\bx_1) & \gamma(\bx_2,\bx_2)
\end{vmatrix}
\label{app:gammaHF}
\end{align}
where $\gamma(\bx,\bx') = \sum_{k=1}^{N} \varphi_k^*(\bx) \varphi_k(\bx')$ is the 1-body density matrix. 
For a Slater determinant wave function, we can thus write the 2-RDM in Eq.~(\ref{app:gammaHF}) as
\begin{align}
\begin{split}
\widetilde{\Gamma}(\bx_1,\bx_2) &= \frac{1}{2} \left\{ \left[\, \sum_{k=1}^{N} |\varphi_k(\bx_1)|^2  \right] \left[\,  \sum_{k'=1}^{N} |\varphi_{k'}(\bx_2)|^2  \right] \right. \\
                                 &\left. - \left[\, \sum_{k=1}^{N} \varphi_k^*(\bx_1) \varphi_k(\bx_2) \right] \left[\, \sum_{k'=1}^{N} \varphi_{k'}^* (\bx_2) \varphi_{k'}(\bx_1) \right] \right\}.
\end{split}
\end{align}
\indent
In our work, we focus on the contribution to exchange from an electron pair occupying two given spin-orbitals, $\varphi_{i\sigma}$ and $\varphi_{j\sigma'}$.  
We thus keep in the 2-RDM only the terms that involve these orbitals, and obtain the density matrix:
\begin{align}
\begin{split}
\!\!\!\!\Gamma(\bx_1,\bx_2) &\!=\! \frac{1}{2} \left\{ \left[ |\varphi_{i\sigma}(\bx_1)|^2 |\varphi_{j\sigma'}(\bx_2)|^2 + |\varphi_{j\sigma'}(\bx_1)|^2 |\varphi_{i\sigma}(\bx_2)|^2 \right] \right. \\
                                 &\!\!\!\!\!\!\!\left. - \left[  \varphi^*_{j\sigma'}(\bx_1) \varphi_{i\sigma}(\bx_1)  \varphi_{j\sigma'}(\bx_2) \varphi^*_{i\sigma}(\bx_2)  + \hc \right] \right\},
\label{app:gamma-HF}                             
\end{split}
\end{align} 
where $\hc$ denotes the Hermitian conjugate, and we removed the tilde to indicate we are no longer working with the full 2-RDM for $N$ electrons, but rather, 
with its two-electron part of relevance here. 
We write explicitly the spatial and spin parts, for each spin-orbital $\varphi_{i\sigma}(\bx)$ and its Hermitian conjugate $\varphi^*_{i\sigma}(\bx)$, as
\begin{align}
\begin{split}
\varphi_{i\sigma} (\bx) &= \varphi_i(\br) \ket{\sigma} \\
\varphi_{i\sigma}^* (\bx) &= \varphi^*_i(\br) \!\bra{\sigma}.
\end{split}
\end{align}
%
%
After substituting in Eq.~(\ref{app:gamma-HF}), we obtain
\begin{align}
\begin{split}
\Gamma_{\sigma\sigma'}(\bx_1,\bx_2) &= \frac{1}{2}\, \big\lvert\, \varphi_i(\br_1)\varphi_j(\br_2) \!\ket{\sigma\sigma'} -  \varphi_j(\br_1)\varphi_i(\br_2) \!\ket{\sigma' \sigma} \big\rvert^2 \\
                                    &=\frac{1}{2} \left\{ \left[\, |\varphi_{i}(\br_1)|^2 |\varphi_{j}(\br_2)|^2 \pr{\sigma \sigma'}{\sigma \sigma'} \right. \right. \\
                                    &\,\,\,\,\,\,\,\,\,\,\,\, \left. +\, |\varphi_{j}(\br_1)|^2 |\varphi_{i}(\br_2)|^2 \pr{\sigma' \sigma}{\sigma' \sigma} \,\right] \\
                                 & \!\!\!\!\!\!\!\! \left. - \left[  \varphi^*_{j}(\br_1) \varphi_{i}(\br_1)  \varphi_{j}(\br_2) \varphi^*_{i}(\br_2)  \pr{\sigma \sigma'}{\sigma' \sigma} + \hc \right] \right\}. 
\label{app:gamma-spin}                             
\end{split}
\end{align}
For given electron spins $\sigma$ and $\sigma'$, the density matrices $\Gamma_{\sigma\sigma'}$ correspond to the configurations in Fig.~\ref{fig1}(c), 
of which we keep only the part contributing to exchange between the orbital $\varphi_i$ and $\varphi_j$.  
For same-spin electrons, $\Gamma_{\sigma\sigma'}$ takes a particularly simple form:
\begin{align}
\begin{split}
\Gamma_{\up\up}(\bx_1,\bx_2) &= |A_{ij}(\br_1,\br_2)|^2 \, \pr{00}{00} \\
\Gamma_{\dn\dn}(\bx_1,\bx_2) &= |A_{ij}(\br_1,\br_2)|^2 \, \pr{11}{11},
\end{split}
\end{align} 
where $A_{ij}(\br_1,\br_2) \!=\! [ \varphi_i(\br_1) \varphi_j(\br_2) - \varphi_j(\br_1) \varphi_i(\br_2) ]/\sqrt{2}$ is the antisymmetric spatial wave function defined above. 
We also see that 
$\Gamma_{\up\up} + \Gamma_{\dn\dn} = \pr{\Psi_1}{\Psi_1} + \pr{\Psi_2}{\Psi_2}$. 
The opposite-spin 2-RDMs can be combined to give:
\begin{align}
\begin{split}
 (\Gamma_{\up\dn} + \Gamma_{\dn\up})&(\bx_1,\bx_2) =  \\
                                                                           & \frac{1}{2} \!\left[\, \big\lvert\, \varphi_i(\br_1)\varphi_j(\br_2) \!\ket{01} -  \varphi_j(\br_1)\varphi_i(\br_2) \!\ket{10} \big\rvert^2 \right. \\
                                                                                     & \left. + \big\lvert\, \varphi_i(\br_1)\varphi_j(\br_2) \ket{10} -  \varphi_j(\br_1)\varphi_i(\br_2) \ket{01} \big\rvert^2 \,\right], 
\end{split} 
\end{align}
so that $\Gamma_{\up\dn} + \Gamma_{\dn\up} =  \pr{\Psi_3}{\Psi_3} + \pr{\Psi_4}{\Psi_4}$. 
Using these relations, the ensemble density matrix $\rho^{(2)}$ for model 2 defined in the main text can be rewritten as 
\begin{equation}
\rho^{(2)}(\bx_1,\bx_2) = \frac{1}{4} \left[ \Gamma_{\up\up} + \Gamma_{\dn\dn} + \Gamma_{\up\dn} + \Gamma_{\dn\up}  \right].
\end{equation}
We have thus shown that $\rho^{(2)}$ is $N$-representable, since it can be expressed as an equal-weight ensemble of 2-RDMs for the four spin configurations in Fig.~\ref{fig1}(c).\\
\indent
The density matrix $\rho^{(1)}$ for the pure state employed in model 1 can be obtained with a similar approach. 
We use the configuration in Fig.~\ref{fig1}(b) with the electrons in orbitals $\varphi_i$ and $\varphi_j$ in the superposition spin state $\ket{s} = (\ket{0} + \ket{1})/\sqrt{2}$.  
Using Eq.~(\ref{app:gamma-spin}) and putting $\ket{\sigma} = \ket{\sigma'} = \ket{s}$, we obtain the 2-RDM
\begin{align}
\begin{split}
\Gamma_{ss}(\bx_1,\bx_2) &= |A_{ij}(\br_1,\br_2)|^2 \pr{ss}{ss} \\
            &= |A_{ij}(\br_1,\br_2)|^2 \times \frac{1}{4} \sum_{\sigma\sigma'\sigma''\sigma'''}^{\{0,1\}} \pr{\sigma\sigma'}{\sigma''\sigma'''}.
\end{split}
\end{align}
Using Eq.~(\ref{def:rho1}), $\rho^{(1)} = \frac{1}{4}\,\sum_{\mu,\nu=1}^4 \pr{\Psi_\mu}{\Psi_{\nu}}$, together with the definitions of the states $\ket{\Psi_\mu}$ in Eq.~(\ref{basis}), 
one can show easily that $\rho^{(1)} = \Gamma_{ss}$. 
We conclude that $\rho^{(1)}$ derives from the 2-RDM for the configuration in Fig.~\ref{fig1}(b).
\vspace{5pt}
\\
\noindent
\textbf{Exchange matrix derivation.} 
Let us discuss briefly how the exchange interaction matrix in Eq.~(\ref{mmatrix}) is obtained. 
The non-zero diagonal matrix elements are $M_{11} \!=\! M_{22} \!=\! J$, while $M_{33} = M_{44} = 0$ since for these matrix elements the exchange integral vanishes. 
We take the orbitals $\varphi_i$ and $\varphi_j$ to be real, so that $M$ is symmetric, and obtain the off-diagonal part of $M$ using the properties of the matrix elements of the two-body operator 
$V_{12}$ between Slater determinants that differ by one or two spin-orbitals (see Ref.~\onlinecite{Grosso}). 
For example, $\!M_{12} \!\!=\!\! \braket{\up\up | V_{12} | \dn\dn}_{\rm X} $ is a matrix element taken between determinants differing by two spin-orbitals; 
it vanishes since no term survives the integration over spin. 
All of the $M_{13}$, $M_{14}$, $M_{23}$, $M_{24}$ also vanish since they are matrix elements between two determinants differing by one spin-orbital, which can be obtained from one another by flipping one spin.
\vspace{5pt}
\\
\noindent
\textbf{Exchange mixing in the entire electron gas.} 
There is an important detail in comparing Eq.~(\ref{pair}) and Eq.~(\ref{final}). 
In the former, $E_{\rm X}^{\rm hybrid}$ is computed using the electron density for doubly-occupied orbitals $\varphi_i$ and $\varphi_j$, which is due to four electrons, while in Eq.~(\ref{final}) the electron density integrates to two electrons. 
This subtlety can be resolved by observing that for the spin-flip exchange interaction to occur, only two electrons (one per orbital) can occupy the orbitals $\varphi_i$ and $\varphi_j$, 
since the initial and final states possess opposite spins in each orbital [see Fig.~\ref{fig3}].  
Therefore, to extend Eq.~(\ref{final}) to four electrons, one needs to use the density $n(\br) = 2\, (\, |\varphi_i(\br)|^2 + |\varphi_j(\br)|^2\,)$ for four electrons in the $3/4\, E_{\rm X}^{\rm DFT}[n]$ term, as in Eq.~(\ref{pair}), while keeping the spin-flip interaction as the one due to two electrons, which equals $1/4\, E_{\rm X}^{\rm HF}$, since this term derives from doubly-excited configurations. 
After this step, our result in Eq.~(\ref{final}) matches exactly the hybrid DFT result in Eq.~(\ref{pair}).  
This reasoning provides the basis for extending the result to the entire electron gas. 
Using the density due to all occupied orbitals, and summing over all orbital pairs the pairwise part accounting for the spin-flip (exact) exchange, 
one obtains the total exchange energy for all the electrons in a hybrid functional with 1/4 mixing, $E_{\rm X} = 3/4\,E_{\rm X}^{\rm DFT} [n]+ 1/4\,E_{\rm X}^{\rm HF}$. 

\end{document}